\documentclass[twocolumn,prl,showpacs,amsmath,amssymb]{revtex4}

\usepackage{graphicx}
\usepackage{bm}

\newcommand{\Fig}[1]{Fig.\;\ref{#1}}

\renewcommand{\propto}{\sim}

\newcommand{\ii}{\ensuremath{\mathrm i}}
\renewcommand{\propto}{\ensuremath{\sim}}

\newcommand{\idt}{\ensuremath{\ii\hbar\,\frac\partial{\partial t}\,}}

\newcommand{\op}[1]{\ensuremath{\hat{#1}}}
\newcommand{\Ham}{\ensuremath{\op H}}

\begin{document}

\preprint{APS/123-QED}

\title{Characterization of strong light-matter coupling in semiconductor quantum-dot microcavities via photon-statistics spectroscopy}

\author{L. Schneebeli}
\email{lukas.schneebeli@physik.uni-marburg.de}
\author{M. Kira}
\author{S.W. Koch}

\affiliation{
Department of Physics and Material Sciences Center,
Philipps-University, 35032 Marburg, Germany 
}

\date{\today}

\begin{abstract}
It is shown that spectrally resolved photon-statistics measurements of the resonance fluorescence 
from realistic semiconductor quantum-dot systems allow for high contrast identification of  
the two-photon strong-coupling states. Using a microscopic theory, the second-rung resonance is analyzed and optimum excitation conditions are determined. The computed photon-statistics spectrum displays gigantic, experimentally robust resonances at the energetic positions of the second-rung emission.
\end{abstract}


\maketitle

Even though light-quantization effects, such as squeezing \cite{Slusher1985}, antibunching \cite{Kimble1977}, and entanglement \cite{Hagley1997}, have been observed under a wide range of conditions, one can rarely detect the discrete energy levels of the different photon-number states directly as resonances. Under strong-coupling conditions between a two-level resonance at the energy $\hbar \omega$ and the quantized resonant light field, the resulting dressed-state resonances appear at $\hbar \omega n \pm g\sqrt{n+1}$ where $n$ is the occupation of the photon-number state $|n \rangle $ and $g$ is the light-matter coupling constant \cite{JC1963}. The resulting discrete energy structure resembles a ladder with $n$-dependent rungs. The first rung shows the so-called vacuum Rabi splitting $2g$, the second rung shows the splitting $2g\sqrt{2}$, and so on. For sufficiently small damping and dephasing, the different rungs are clearly visible as discrete resonances, e.g., in transmission spectra. 

The first observations of strong coupling \cite{Thompson1992,Boca2004} and the second rung \cite{Brune1996} were reported for atoms in high-quality cavities. Besides the demonstration of the discrete nature of light, this research has lead to major advances, e.g. in utilizing entanglement effects as a basis for quantum-logic applications \cite{Wallraff2004,Imamoglu1999}.
The clear identification of the second-rung resonance can be considered as the critical step if one wants to use other materials, such as solid-state systems for strong-coupling applications. A promising candidate are semiconductor quantum dots (QD) in microcavities. Recent experiments \cite{Reithmaier2004,Yoshie2004,Peter2005,Hennessy2007} have already demonstrated that one can see clear vacuum Rabi splitting effects in the photoluminescence (PL) of such systems. However, despite continuing attempts, the second rung resonance has not yet been observed. The most likely reason for this failure is that dephasing and the other broadening effects in the real systems smear out the expected discrete features.

In this Letter, we propose a novel scheme to observe the second-rung strong coupling effects even under the
realistic broadening conditions of the currently available samples. In our approach, we use a fully 
microscopic theory to determine the exact conditions to obtain unambiguous evidence for the presence of the second-rung resonance in QD microcavities. Our results confirm the difficulty of the second-rung
observations in standard semiconductor experiments that monitor PL after carrier capture of electrons, initially created to the wetting layer. However, we demonstrate that it should
be completely feasible to observe the second rung in resonance fluorescence measurements where the QD-cavity system is resonantly laser pumped while the re-emitted light spectrum is recorded. Our results 
clearly show that the spectrum of the two-photon correlations, $g^{(2)}$, contains a gigantic resonance at the energetic position of the second rung. Due to its large value, this $g^{(2)}$ resonance remains visible even when scattering becomes appreciable making this photon-statistics spectroscopy a viable experimental scheme to detect the second rung in realistic QD systems.

The first QD strong-coupling experiments \cite{Reithmaier2004,Yoshie2004,Peter2005} were performed with QDs grown on top of a quantum well (QW) which acts as the so-called wetting layer (WL). Due to the three-dimensional confinement, the QD electrons occupy a discrete set of states which couple to each other and to a continuum of WL states via the Coulomb, phonon, and light-matter interactions, yielding a complicated many-body problem \cite{Brasken2000,Baer2004,Wojs1996,Feldtmann2006}. For strong-coupling investigations, one typically studies strongly confining dots where only one discrete electron and hole level -- constituting an effective two-level system -- couples to the cavity resonance. This scenario simplifies the analysis considerably because the remaining WL electrons and the phonon interaction effectively act as noise reservoirs to the isolated two-level system. As in the real systems, we assume that a variable number of dots is positioned inside the cavity. The different dots are labeled by $j$ and the corresponding electrons and holes are described via Fermionic operators $\hat{e}_j$, $\hat{h}_j$, with eigen energy $E^e$, $E^h$, respectively. The light field corresponding to a mode function $u_q({\bf r})$ is quantized by introducing the Bosonic operator $\hat{B}_q$ where $q$ is the wave vector of light outside the cavity. The system Hamiltonian for the investigated dots and its interaction with light follows then from  \cite{PQE2006,Feldtmann2006}
$\Ham = \sum_{j}  \left( E^e \hat{e}_j^\dagger
\hat{e}_j +  E^h \hat{h}_j^\dagger \hat{h}_j \right) + \sum_q \hbar \omega_q \left(\hat{B}_q^\dagger \hat{B}_q  + \frac 12\right) +\sum_{q j} \left({\cal F}_q^{\star} \hat{B}_q^\dagger \hat{h}_j \hat{e}_j + \mathrm{h.c.}\right)$,
%
where $\hbar \omega_q = \hbar |q| c$ is the photon energy and ${\cal F}_q = -i d {\cal E}_q u_q({\bf r}_j)$ defines the coupling strength via the dipole-matrix element $d$, the vacuum-field amplitude ${\cal E}_q$, and the position of the dot ${\bf r}_j$. The used multi-mode description of the light field allows us to flexibly describe the coupling between an external pump field to the internal cavity field, propagation effects, as well as the finite linewidth of the cavity mode without additional phenomenological parameters. To get a physically correct form for the light modes, we solve $u_q({\bf r})$ for a cavity between distributed Bragg reflector (DBR) mirrors. Since the resulting $|u_q({\bf r})|^2$ produces nearly a Lorentzian resonance \cite{PQE1999}, this model is well-suited for other types of cavities after we adjust the cavity resonance $\hbar \omega_c$ and its half width $\gamma_\mathrm{cav}$ to match the studied experimental cavity. In this connection, it is convenient to determine the quality factor $Q=\hbar \omega_c/(2\gamma_\mathrm{cav})$. Furthermore, the vacuum Rabi splitting $2g=2d{\cal E}_c \sqrt{N_\mathrm{dot} \sum_q |u_q({\bf r}_j)|^2}$ can also be adjusted to the experiment having $N_\mathrm{dot}$ dots within the cavity. 

Since $\hat{H}$ involves infinitely many light modes, it is not feasible to solve the corresponding QD emission by computing the wave-function or density-matrix dynamics. Therefore, we evaluate the time-evolution of the relevant expectation values directly. Due to the quantum-optical light-matter coupling within $\hat{H}$, the corresponding Heisenberg equations of motion formally produce an infinite hierarchy of equations. We systematically truncate this hierarchy with the so-called cluster-expansion approach \cite{Wyld1963,PQE2006}. At the lowest level, we find the usual Maxwell-Bloch equations \cite{PQE2006, Allen1987} for coherent light  $ \langle  \hat{B}_q  \rangle $, QD polarization $P_j =  \langle  \hat{h}_j \hat{e}_j  \rangle $, as well as electron (hole) occupations $f_j^e =  \langle  \hat e^\dagger_j \hat{e}_j  \rangle $ ($f_j^h =  \langle \hat h^\dagger_j \hat{h}_j  \rangle $). We assume that the pump pulse is a classical field defined by $ \langle  \hat E (r)  \rangle  = \sum_q i {\cal E}_q u_q (r)  \langle  \hat{B}_q  \rangle  +{\rm c.c. }$ positioned initially outside the cavity. As it propagates inside the cavity it generates the QD polarization and densities. 

At the same time, the light-matter coupling induces quantum-optical correlations. In particular, we concentrate on the analysis of squeezing generation described by the dynamics of the two-photon correlations $\Delta  \langle \hat{B}_q \hat{B}_{q'} \rangle \equiv  \langle \hat{B}_q \hat{B}_{q'} \rangle  -  \langle \hat{B}_q \rangle  \langle \hat{B}_{q'} \rangle $ following from
\begin{eqnarray}
\label{eq:BB-dynamics}
\idt \Delta \langle \hat{B}_q \hat{B}_{q'} \rangle 
  &=& \hbar \left( \omega_q + \omega_{q'} \right)
  \Delta \langle  \hat{B}_q \hat{B}_{q'}  \rangle 
\\\nonumber
   &+&  
   \sum_j 
   	 \Delta \langle 
   	\left(
   	{\cal F}_{q'}^{\star} \hat{B}_q
   	+
   	{\cal F}_q^{\star}
   	\hat{B}_{q'} 
   	\right)
   	 \hat{P}_j  \rangle,
\\
\label{eq:BP-dynamics}
   \idt \Delta \langle \hat{B}_q \hat{P}_j  \rangle
    &=&
    \left( \hbar \omega_q + E^{eh} -i \gamma_P \right)
		\Delta \langle \hat{B}_q \hat{P}_j \rangle
\\\nonumber
   &+& \left(1-f^e_j-f^h_j \right) \sum_{q'} {\cal F}_{q'} \Delta \langle \hat{B}_q \hat{B}_{q'} \rangle   
\\\nonumber
  &+& \Omega_j \, \Delta \langle     
  \hat{B}_q \hat{f}^{eh}_j   \rangle  
  - {\cal F}_q^{\star}  P^2_j 
  + T\left[\Delta \langle 3 \rangle \right].
\end{eqnarray}
Here we have defined $E^{eh} \equiv E^e+E^h$, the polarization operator $\hat{P}_j \equiv \hat{h}_j \hat{e}_j$, the density operator $\hat{f}^{eh}_j = \hat e^\dagger_j \hat{e}_j + \hat h^\dagger_j \hat{h}_j$, as well as the classical Rabi energy $\Omega_j = d \langle \hat{E}({\bf r}_j) \rangle$ at the QD position. The coupling to the WL continuum and the phonons introduces dephasing for polarization-dependent quantities, included phenomenologically via the dephasing constant $\gamma_P$. Otherwise, we explicitly include effects up to three-particle correlations, denoted as $\Delta \langle 3 \rangle$. We observe that the squeezing signal $\Delta \langle \hat{B} \hat{B} \rangle$ couples to the correlated destruction of both a photon and an electron-hole pair, $\Delta \langle \hat{B} \hat{P} \rangle$. This correlation is created spontaneously via the nonlinear polarization source $P^2$. After its creation,  $\Delta \langle \hat{B} \hat{P} \rangle$ produces non-vanishing $\Delta \langle \hat{B} \hat{B} \rangle$. The generated squeezing signatures are coupled back to $\Delta \langle \hat{B} \hat{P} \rangle$ via the stimulated contribution $\sum \Delta \langle \hat{B} \hat{B} \rangle$ in Eq.~(\ref{eq:BP-dynamics}), which together with
the photon-density correlations, $\Delta \langle \hat{B} \hat{f}^{eh} \rangle$, eventually produce the different rungs. Equations (\ref{eq:BB-dynamics})--(\ref{eq:BP-dynamics}) are structurally similar to those obtained for $\Delta \langle \hat{B}_q^\dagger \hat{B}_{q'} \rangle$ and $\Delta \langle \hat{B}_q^\dagger \hat{P}_j \rangle$.

We evaluate our theory using the parameters of the three different, recently published experimental configurations in which vacuum Rabi splitting has been reported. The QD-pillar investigations \cite{Reithmaier2004,Loeffler2005} have $n_{\rm dot} = 1.3\cdot 10^9 \, \mathrm{cm}^{-2}$ within DBR mirrors yielding a quality factor of $Q=2.4 \cdot 10^4$ with cavity frequency $\hbar \omega_c = 1.33$ eV. The effective cavity area is $S = 3.0 \, \mathrm{\mu m}^2$ yielding 
$N_{\rm dot} = 39$ and $g=20$ GHz. In another QD-crystal experiment \cite{Yoshie2004},  $n_{\rm dot} = 6.0\cdot 10^9 \, \mathrm{cm}^{-2}$ QDs were placed within a photonic crystal providing $Q=2.2 \cdot 10^4$, $\hbar \omega_c = 1.0$ eV, $S = 10\, \mathrm{\mu m}^2$,  $N_{\rm dot} = 600$, and $g=22$ GHz. In the QD-disk example \cite{Srinivasan2007}, $n_{\rm dot} = 10^{10} \, \mathrm{cm}^{-2}$ QDs were positioned within a microdisk giving $\hbar \omega_c = 1.0$ eV, $Q=4\cdot 10^5$, $S = 2.5\, \mathrm{\mu m}^2$, $N_{\rm dot} = 250$, and $g=11$ GHz. The dot dipole moment is $d=5.3$ \AA e in all these systems.

In all cases, we carefully match the system parameters with our microscopic model. As an initial condition for the resonance-fluorescence computations, we assume that the QDs are initially unexcited. We then solve the full set of equations including the three-particle correlations to determine the re-emission following the excitation. Especially, we monitor the photoluminescence spectrum $I(\omega_q) \equiv \langle \hat B^\dagger_q \hat B_q \rangle$ in directions different than the excitation together with the two-photon correlation spectrum $ g^{(2)} (\omega_q) \equiv  \langle \hat{B}_q^\dagger \hat{B}_q^\dagger \hat{B}_q \hat{B}_q \rangle / \langle \hat{B}_q^\dagger \hat{B}_q \rangle ^2$,
determining the probability of detecting two photons with momentum $q$ at the same time. We have verified with an independent single-mode analysis that the four-photon quantity can be accurately described by its singlet-doublet-triplet truncation.

At the level of the Jaynes-Cummings model \cite{JC1963}, the second-rung wave function follows from 
$| \phi_\pm \rangle \propto |1\rangle | {\rm up} \rangle \pm |2\rangle | {\rm down}  \rangle$ where $|{\rm up} \rangle$ ($|{\rm down} \rangle$) refer to the excited (unexcited) dot/atom.
Thus, one can access this state either by having the system originally in the excited state and providing sufficient occupation of the $|1\rangle $ photon state or when the system is originally unexcited and the light has a strong occupation of the state  $|2\rangle$. If carriers are created nonresonantly to the wetting layer, the carrier-capture processes will bring the dot toward the excited state $|{\rm up} \rangle$ with a characteristic time $\tau_\mathrm{capt}$. At the same time, the excited dot emits light into the cavity mode. A typical time scale to create $| 1 \rangle$ out of the vacuum is defined by half of the Rabi period, giving $\tau_\mathrm{Rabi} \equiv \frac{\hbar}{g}$. Thus, the carrier capture has to support both the dot and cavity excitation, which decays with the lifetime $\tau_\mathrm{cav}$. To clearly see the second-rung from PL alone, one thus must have $\tau_\mathrm{capt} \ll \tau_\mathrm{Rabi}$ and  $\tau_\mathrm{capt} \ll \tau_\mathrm{cav}$, which is difficult to fulfill
in current experiments since the typical capture times of $\tau_\mathrm{capt} = 50$ ps \cite{Ohnesorge1996} exceed both the Rabi ($11$ ps) and the cavity lifetime ($25$ ps) \cite{Yoshie2004}. Further problematic aspects are the dephasing and the incomplete carrier capture resulting in $f^{e,h} <1$.
%
\begin{figure}
\centerline{\scalebox{0.85}{\includegraphics{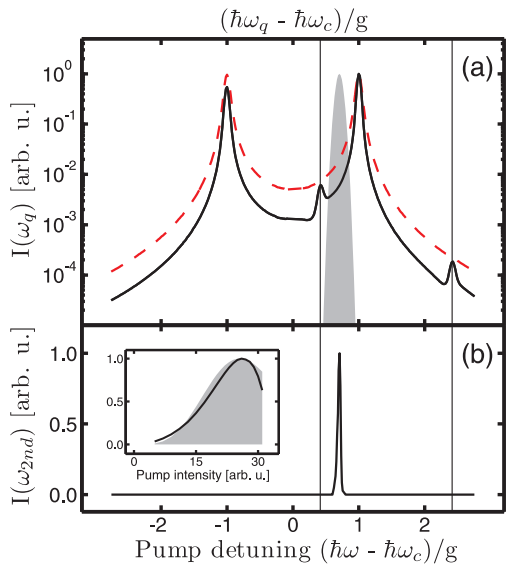}}}
\caption{Resonance fluorescence spectra for a QD-disk system $(\Delta = 0)$. (a) Spectrum of the resonant second-rung pump (shaded area) and resulting emission for low dephasing $\gamma_P = 0.06$ GHz (solid line) and elevated dephasing $\gamma_P=0.4$ GHz (dashed line). The vertical lines mark the second-rung resonances. (b) The second-rung emission intensity as function of pump frequency. The inset shows the second-rung emission intensity and the two-photon state occupation $P_2$ (shaded area) in the pump field as function of pump intensity.} 
\label{PLemOpt}
\end{figure}

The second-rung state can also be reached by bringing the cavity directly into the Fock state $|2 \rangle$ while the dot remains unexcited ($| {\rm down} \rangle$). A resonant excitation of the cavity-dot system with a coherent laser, described by $| \alpha \rangle = \sum_{n=0}^\infty \alpha^n/\sqrt{n!} \exp\left[-|\alpha|^2/2\right]|n \rangle$, always produces some occupation  at $| 2 \rangle$. In particular, the $| 2 \rangle$ component of the external light is selectively converted to the second-rung state $| \phi_+ \rangle$ if its energy $E_\mathrm{pump} = 2 \hbar \omega$ matches the energy of the dressed dot-cavity state $E_{\rm dress} = 2 \hbar \omega_c + \sqrt{2} g$, giving the condition $\hbar \omega = \hbar \omega_c +  \frac{g}{\sqrt{2}}$ for zero cavity-dot detuning $\Delta\equiv E^{eh} - \hbar\omega_c$. The transfer of other $|n\rangle$ states is suppressed. For non-zero $\Delta$, the optimum energy is defined by 
$\hbar \omega = \hbar \omega_c + (\Delta +\sqrt{ \Delta^2 + 8 g^2})/4$.
This condition guarantees the selective excitation at the second rung with a probability determined by the two-photon state occupation $P_2=\frac{|\alpha|^4}{2} \; e^{-|\alpha|^2}$ within the external pump.
We also notice that the second-rung emission energy, $\hbar \omega_{\rm 2nd} = \hbar \omega_c + (\sqrt{\Delta^2+8g^2}\pm\sqrt{\Delta^2+4g^2})/2$, and the optimum pump energy $\hbar \omega$ are generally nondegenerate.

To illustrate how the resonant second-rung pumping works, we consider the excitation of the QD-disk system ($\Delta =0$) with coherent light energetically centered at $\hbar \omega = \hbar \omega_c +  \frac{g}{\sqrt{2}}$. Figure \ref{PLemOpt}(a) shows the pump spectrum (shaded area) together with the resulting emission $I(\omega)$ for the low $\gamma_P = 0.06$ GHz dephasing (solid line) and the elevated $\gamma_P = 0.4$ GHz dephasing (dashed line). Besides the usual vacuum-Rabi peaks at $\hbar \omega_c \pm g$ we observe for low $\gamma_P$ that the fluorescent light contains clear second-rung emission peaks at $\hbar \omega_\mathrm{2nd} = \hbar \omega_c + \left(\sqrt{2}\pm1\right) g$. This features are clear signatures of strong coupling even though they gradually vanish for increasing $\gamma_P$.
If the same analysis is repeated for non-resonant excitation, $I(\omega)$ shows just the vacuum-Rabi peaks, regardless of $\gamma_P$. Thus, it is clearly easier to demonstrate true strong coupling using the resonant second-rung pumping. To verify that the emerging resonances result from the second rung, we performed additional test by changing the frequency and
intensity of the pump. Figure \ref{PLemOpt}(b) shows $I_{\rm 2nd} \equiv I(\omega_{\rm 2nd})$ at the lower second-rung peak as function of pumping $\hbar \omega$. Indeed, $I_{\rm 2nd}$ peaks exactly at the optimum pumping frequency $\hbar \omega = \hbar \omega_c + g/\sqrt{2}$. Furthermore, the inset shows that $I_{\rm 2nd}$  scales with the two-photon state occupation $P_2$ in the pump field, as the coherence parameter $\alpha$ is increased. Thus, it is very important to properly adjust not only the excitation frequency but also the excitation intensity such that the two-photon state is sufficiently occupied. 
%
\begin{figure}
\leftline{\scalebox{1.0}{\includegraphics{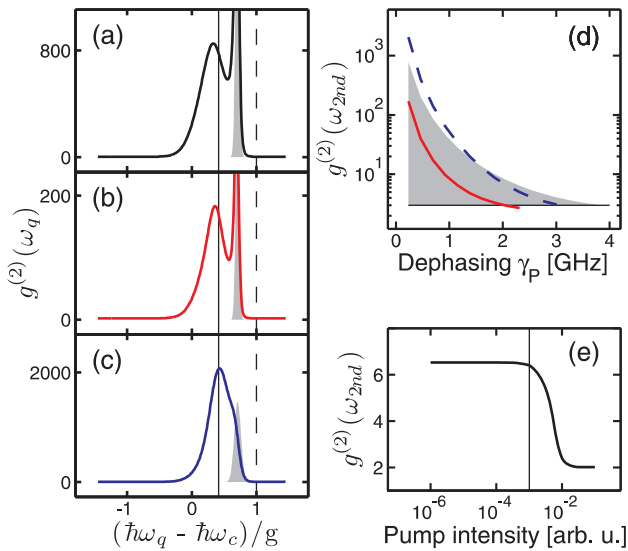}}}
\caption{Photon-statistics spectra (solid line) for (a) QD-pillar, (b) QD-crystal, and (c) QD-disk system, at the time when the pump (shaded area) has its maximum. In all systems, $\Delta = 0$ and $\gamma_P = 0.23$ GHz. (d) The corresponding $g^{(2)}$ at the second rung as function of dephasing for the QD-pillar (shaded), QD-crystal (solid), and QD-disk (dashed) systems. The horizontal line at $g^{(2)}=3$ indicates the visibility limit. (e) The pump-intensity dependence of $g^{(2)}$ at the second rung in QD-pillar system ($\gamma_p = 2.3$ GHz). The vertical line marks the applied pump intensity for (a)-(d).} \label{g2Time}
\end{figure}

Even though the second-rung pumping induces true strong coupling effects in the QD system, the realistic dephasing of $\gamma_P=0.4$ GHz washes out the most intriguing features in the standard experiments. Thus, we analyze next the $g^{(2)}$ correlations and show that they can serve as more robust signatures. In \Fig{g2Time}(a)-(c), the solid line presents the computed $g^{(2)}$ spectrum for the different QD-cavity systems after resonant second-rung pumping (shaded area). The energetic position of the second rung (upper NMC peak) is marked by the solid (dashed) vertical line. Our results verify that all QD-cavity systems yield $g^{(2)}$ resonances with gigantic values close to $10^3$ at the second-rung energy. 

This strongly enhanced $g^{(2)}$ follows from the fundamental properties of the resonant second-rung pumping which exclusively enables the Fock-state $|2\rangle$ to interact with the QD. Since the cavity initially is in the vacuum state, the addition of this Fock state essentially creates cavity light into state $| 0 \rangle + {\sqrt P_2} | 2 \rangle$, which is a squeezed state with an appreciably small $P_2$. The same conclusion follows from Eq.~(\ref{eq:BB-dynamics}) showing that the squeezing correlations $\Delta \langle \hat{B} \hat{B} \rangle$ are created in this process. It is well-known \cite{Mattinson2002} that a squeezed state close to a vacuum produces a very large $g^{(2)}$ when it interacts with a Fermionic system. Thus, the gigantic $g^{(2)}$ seems to be directly connected with the squeezing generated by the resonant second-rung pumping. 

Since the $g^{(2)}$ resonance is very large for small $\gamma_P$ it remains clearly visible even for elevated $\gamma_P$. To show this, \Fig{g2Time}(d) presents for the three different QD-cavity systems the computed value of $g^{(2)}$ at the spectral position of the second rung as function of dephasing $\gamma_P$. The horizontal line at $g^{(2)}=3$ serves as a visibility limit for the observability of the second rung. We see that for all cases, a clear resonance occurs even for dephasing values as large as $\gamma_P = 2$ GHz. This is considerably larger than the natural dot dephasing of $\gamma_P = 0.3$ GHz \cite{Bayer2002} and it is in the range of the broad cavity widths $\gamma_\mathrm{cav}=5-6$ GHz of the QD-pillar and QD-crystal cavity. From \Fig{g2Time}(a)-(c), we also can  see that there is no vacuum-Rabi peak in the $g^{(2)}$ spectrum. Thus, the $g^{(2)}$ spectroscopy provides a unique resonance at the second-rung position. Since the pump and the second-rung energies are different, the squeezing-generated  $g^{(2)}$  feature around the pumping energy can always be distinguished from the actual second-rung peak. 
Figure \Fig{g2Time}(e) presents the $g^{(2)}$ second-rung resonance as function of pump intensity for a large dephasing $\gamma_p=2.3$ GHz. We notice that the $g^{(2)}$ signal remains unchanged in the low-intensity regime but decreases for too strong excitation. Thus, the resonant second-rung pumping has to be performed in the low-intensity regime where $g^{(2)}(\omega_\mathrm{2nd})$ approaches a constant value. The calculations in \Fig{g2Time}(a)-(d) were done in this stable regime, as indicated by the vertical line in \Fig{g2Time}(e).

In summary, our proposed method of photon-statistics spectroscopy identifies a way for the experimental verification of strong-coupling resonances in the resonance fluorescence of semiconductor QD microcavities. Based on a consistent microscopic analysis, we predict that especially the two-photon strong-coupling state 
should be clearly visible as a pronounced resonance in the $g^{(2)}$ spectrum. These measurements allow for clear identification of true strong coupling situations in realistic QD systems.

Acknowledgements: This work is supported by the Quantum Optics in Semiconductors DFG Research Group. We thank H.M. Gibbs and G. Khitrova for valuable discussions.


\end{document}